\documentclass{kluwer}
\usepackage{graphicx}

\newcommand{\citeN}[1]{\citeauthor{#1}\ (\citeyear{#1})}
\newcommand{\citeNP}[1]{\citeauthor{#1},\ \citeyear{#1}}

\newcommand{\apj}{{\it Astrophys.~J.}}

\newcommand{\aap}{{\it Astron.~Astrophys.}}

\newcommand{\an}{{\it Astr.~Nachr.}}

\newcommand{\mnras}{{\it Monthly Notices Roy.\ Astron.\ Soc.}}

\newcommand{\solphys}{{\it Solar~Phys.}}

\newcommand{\ujlisti}{
\itemsep=0 em
\parsep=0.5 em
\partopsep=0.25 em
\topsep=0 em}
\newcommand{\ujlistii}{
\itemsep=0 em
\parsep=0.5 em
\partopsep=0.25 em
\topsep=0 cm}

\newenvironment{lista}{\begin{list}{--}{\ujlisti}}{\end{list}}

\renewcommand{\[}{\begin{equation}}
\renewcommand{\]}{\end{equation}}
\newcommand{\scri}{\scriptsize}

\newcommand{\ov}{\overline}

\def\la{\mathrel{\mathchoice {\vcenter{\offinterlineskip\halign{\hfil
 $\displaystyle##$\hfil\cr<\cr\sim\cr}}}
 {\vcenter{\offinterlineskip\halign{\hfil$\textstyle##$\hfil\cr
 <\cr\sim\cr}}}
 {\vcenter{\offinterlineskip\halign{\hfil$\scriptstyle##$\hfil\cr
 <\cr\sim\cr}}}
 {\vcenter{\offinterlineskip\halign{\hfil$\scriptscriptstyle##$\hfil\cr
 <\cr\sim\cr}}}}}
\def\ga{\mathrel{\mathchoice {\vcenter{\offinterlineskip\halign{\hfil
 $\displaystyle##$\hfil\cr>\cr\sim\cr}}}
 {\vcenter{\offinterlineskip\halign{\hfil$\textstyle##$\hfil\cr
 >\cr\sim\cr}}}
 {\vcenter{\offinterlineskip\halign{\hfil$\scriptstyle##$\hfil\cr
 >\cr\sim\cr}}}
 {\vcenter{\offinterlineskip\halign{\hfil$\scriptscriptstyle##$\hfil\cr
 >\cr\sim\cr}}}}}

\newcommand{\rcz}{r_{\mbox{\scriptsize bcz}}}
\newcommand{\rin}{r_{\mbox{\scriptsize in}}}
\newcommand{\rim}{r_{\mbox{\scriptsize im}}}

\newcommand{\omegacyc}{\omega_{\mbox{\scriptsize cyc}}}
\newcommand{\Hskin}{H_{\mbox{\scriptsize skin}}}

\begin{document}                                                                                   
\begin{article}
\begin{opening}

\title{TACHOCLINE CONFINEMENT BY AN OSCILLATORY MAGNETIC FIELD}
\subtitle{}

\author{E. \surname{Forg\'acs-Dajka}}
\author{K. \surname{Petrovay}}
\institute{E\"otv\"os University, Dept.~of Astronomy, Budapest, Pf.~32,
	   H-1518 Hungary} 

\date{[{\it Solar Physics}, submitted (2001)]}

\runningtitle{Tachocline Confinement by an Oscillatory Magnetic Field}
\runningauthor{Forg\'acs-Dajka \& Petrovay}

\begin{abstract} 
Helioseismic measurements indicate that the solar tachocline is very thin, its
full thickness not exceeding 4\,\% of the solar radius. The mechanism that
inhibits differential rotation to propagate from the convective zone to deeper
into the radiative zone is not known, though several propositions have been
made. In this paper we demonstrate by numerical models and analytic estimates 
that the tachocline can be confined to its observed thickness by a poloidal
magnetic field $B_p$ of about one kilogauss, penetrating below the convective
zone and oscillating with a period of 22 years, if the tachocline region is
turbulent with a diffusivity of $\eta\sim 10^{10}\,$cm$^2/$s (for a turbulent
magnetic Prandtl number of unity). We also show that a similar confinement may
be produced for other pairs of the parameter values ($B_p$, $\eta$). The
assumption of the dynamo field penetrating into the tachocline is consistent
whenever $\eta\ga 10^9\,$cm$^2/$s. 
\end{abstract}


\end{opening}

\section{Introduction}

Helioseismic inversions of the solar internal rotation during the past decade 
invariably showed that the surface-like latitudinal differential rotation,
pervading the convective zone, changes to a near-rigid rotation in the
radiative zone. The change takes place in a very thin layer known as the
tachocline (e.g.\ \citeNP{Kosovichev:tachocline}). Known properties of the
tachocline were recently reviewed by \citeN{Corbard+:SOGO}. While for a long
time only upper limits were available for the thickness of the tachocline, a
more exact determination of the thickness was recently made by several groups
(\citeNP{Corbard+:AA98}; \citeNP{Corbard+:AA99};
\citeNP{Char+:tacho.thickness}; \citeNP{Basu+Antia:SOGO}). Their findings can
be summarized as follows. Below the equator the tachocline is centered around
at $R=0.691\pm 0.003$ solar radii, and its full thickness is $w=0.04\pm0.014$
solar radii.\footnote{By ``full thickness'' here we mean the radial interval
over which the horizontal differential rotation is reduced by a factor of 100.
Different authors use different definitions of $w$, which explains most of the
variation among various published values. The $e$-folding height of
differential rotation, i.e.\ the scale height of the tachocline is then
$H=w/\ln 100\la 0.01\,R_\odot$.} Compared with the internal radius of the
(adiabatically stratified) convective zone ($\rcz=0.71336 \pm 0.00002$), this
implies that the tachocline lies directly beneath the convective zone. There
seems to be significant evidence for a slightly prolate form of the tachocline
($R=0.71\pm0.003$ solar radii at a latitude of $60^\circ$) and marginal
evidence for a thicker tachocline at high latitudes ($w= 0.05\pm 0.005$ solar
radii at $60^\circ$). This seems to indicate that at higher latitudes up to
half of the tachocline lies in the adiabatically stratified convective zone.

A thorough understanding of the physics of the tachocline is crucial for 
understanding the solar dynamo for several reasons (see
\citeNP{Petrovay:SOLSPA}). Firstly, the shear due to differential rotation is
generally thought to be responsible for the generation of strong toroidal
fields from poloidal fields. This shear is undoubtedly far stronger in the
tachocline than anywhere else in the Sun. Second, both linear and nonlinear
stability analyses of toroidal flux tubes lying at the bottom of the convective
zone show that the storage of these tubes  for times comparable to the solar
cycle is only possible below the unstably stratified part of the convective
zone, in a layer that crudely coincides with the tachocline. Indeed, as flux
emergence calculations now provide ample evidence that solar active regions
originate from the buoyant instability of $10^5\,$G flux tubes lying in the
stably stratified layers below the convective zone proper
(\citeNP{FMI:Freibg}), it is hard to evade the conclusion that strong magnetic
fields oscillating with the dynamo period of 22 years {\it must\/} be present
in (at least part of) the tachocline.

A magnetic field oscillating with a circular frequency $\omegacyc=2 \pi /P$, 
$P= 22$ years is
known to penetrate a conductive medium only down to a skin depth of 
\[ \Hskin=(2\eta/\omegacyc)^{1/2} \label {eq:skin} \]  
(cf.\ \citeNP{Garaud:skin}). Using a molecular value for the magnetic 
diffusivity $\eta$, this turns out to be very small (order of a few kilometers)
in the solar case, apparently suggesting that the oscillating dynamo field
cannot penetrate very deep into the tachocline region. Note, however, that in
order to store a magnetic flux of order $10^{23}\,$Mx in the form of a toroidal
field of $10^5\,$ G in an active belt of width $\sim 10^5\,$km, the storage
region must clearly have a thickness of at least several megameters. (Or
possibly more, taking into  account the strong magnetic flux loss from emerging
loops ---cf.\ \citeNP{Petrovay+FMI:erosion}; Dorch, private communication.)
This is only compatible with equation (\ref{eq:skin}) for $\eta\ga
10^{9}\,$cm$^2/$s. 

A turbulent magnetic diffusivity of this order of magnitude is not implausible,
given our present lack of detailed information about the physical conditions in
the tachocline. The value required is still several orders of magnitude below
the diffusivity in the solar convective zone ($\sim 10^{13}\,$cm$^2/$s). The
turbulence responsible for this diffusivity may be generated either by
non-adiabatic overshooting convection or by MHD instabilities of the tachocline
itself. Note that several recent analyses have treated the problem of the
stability of the tachocline under various approximations (e.g.\
\citeNP{Gilman+Dikpati:unstable.tacho}; see review by
\citeNP{Gilman:tacho.review}). The results are not conclusive yet, but it is
clearly quite possible that, once all relevant (three-dimensional, nonlinear,
MHD) effects are taken into account,  the tachocline will prove to be unstable
and therefore capable of maintaining a certain level of turbulence. Note
that the remaining slight inconsistencies between the standard and seismic
solar models also seem to indicate some extra (probably turbulent) mixing in
the tachocline layer (\citeNP{Gough:seismo.review}). On the other hand, this
turbulent diffusivity should not extend below a depth of a few times 10 Mm to
avoid an overdepletion of lithium. Such a shallow depth of the turbulent layer
is consistent with the seismic constraints on tachocline thickness, quoted
above.

These considerations prompt us to consider the structure of a turbulent
tachocline pervaded by an oscillatory magnetic field. For simplicity, the
poloidal field will be treated as given, in the form of a simple oscillating
field of characteristic strength $B_p$, of dipolar latitude-dependence,
penetrating below the convective zone to a shallow depth of about 30 Mm
($0.04\,R_\odot$). We will find that the observed tachocline thickness is
reproduced for suitable pairs of the parameter values ($B_p$, $\eta$). This
effect would offer a straightforward explanation for the thinness of the
tachocline. Indeed, this thinness implies a horizontal transfer of angular
momentum that is much more effective than the vertical transport. In our model,
the horizontal transport is due to Maxwell stresses in the strong oscillatory
magnetic field.  Alternative mechanisms proposed for the horizontal transport
include a strongly horizontally anisotropic turbulence (\citeNP{Spiegel+Zahn}),
non-diffusive hydrodynamical momentum fluxes (\citeNP{Dajka+Petrovay:SOGO}) as
well as Maxwell stresses due to a weak permanent internal magnetic field in the
radiative interior (\citeNP{McGregor+Char}; \citeNP{Rudiger+Kichat:thin.tacho};
\citeNP{Garaud:SOGO}). The plausibility of the processes implied by the the 
hydrodynamical scenarios is, however, dubious. As pointed out by
\citeN{Canuto:tacho}, the extremely strong horizontal anisotropy necessary to
limit the tachocline to within 4\,\% of the solar radius has never been
observed in nature or in laboratory, while non-diffusive fluxes also require an
unrealistically high amplitude to do the trick. Internal magnetic fields are
certainly able to confine a non-turbulent tachocline (and they will probably be
needed to explain the lack of {\it radial\/} differential rotation in the solar
interior anyway). Nevertheless, on basis of the arguments outlined above we
feel that the alternative concept of a turbulent tachocline pervaded and
confined by a dynamo-generated field is worth considering.

\section{Estimates}

Let us first regard the following model problem. 
Consider a plane parallel layer of incompressible fluid of density $\rho$,
where the viscosity $\nu$ and the magnetic diffusivity $\eta$ are taken to be
constant. At $z=0$ where $z$ is the vertical coordinate (corresponding to depth
in the solar application we have in mind) a periodic horizontal shearing flow
is imposed in the $y$ direction:
\[ v_{y0}= v_0 \cos(kx) \]
(so that $x$ will correspond to heliographic latitude, while $y$ to the
longitude). We assume a two-dimensional flow pattern ($\partial_y=0$) and
$v_x=v_z=0$ (no ``meridional flow''). An oscillatory horizontal ``poloidal''
field is prescribed in the $x$ direction as
\[ B_x=B_p \cos(\omega t) \]
The evolution of the azimuthal components of the velocity and the magnetic
field is then described by the corresponding components of the equations of
motion and induction, respectively. Introducing $v=v_y$ and using Alfv\'en
speed units for the magnetic field
\[ V_p=B_p(4\pi\rho)^{-1/2} \qquad b= B_y(4\pi\rho)^{-1/2} , \]
these can be written as
\[ \partial_t v=V_p \cos(\omega t)\partial_x b-\nu\nabla^2 v  \label{eq:veq} \]
\[ \partial_t b=V_p \cos(\omega t)\partial_x v-\eta\nabla^2 b \label{eq:beq} \]
Solutions may be sought in the form
\[ v=\ov v(x, z) +v'(x,z) f(\omega t) \]
\[ b=b'(x,z) f(\omega t+\phi) \]
where $f$ is a $2\pi$-periodic function of zero mean and of amplitude ${\cal
O}(1)$.
($\ov a$ denotes time average of $a$, while $a'\equiv a-\ov a$.)

It will be of interest to consider the (temporal) average of equation
(\ref{eq:veq}):
\[ 0=V_p \ov{\cos(\omega t)f(\omega t+\phi)} \partial_x b' -\nu\nabla^2 \ov v 
   \label{eq:vmean}  \]
Subtracting this from equation (\ref{eq:veq}) yields
\[ \partial_t v'=V_p [\cos(\omega t)f(\omega t+\phi)]'\partial_x b
    -\nu\nabla^2 v'  \label{eq:vfluc} \]

\begin{figure}[!t]
\centering
\includegraphics[width=8cm]{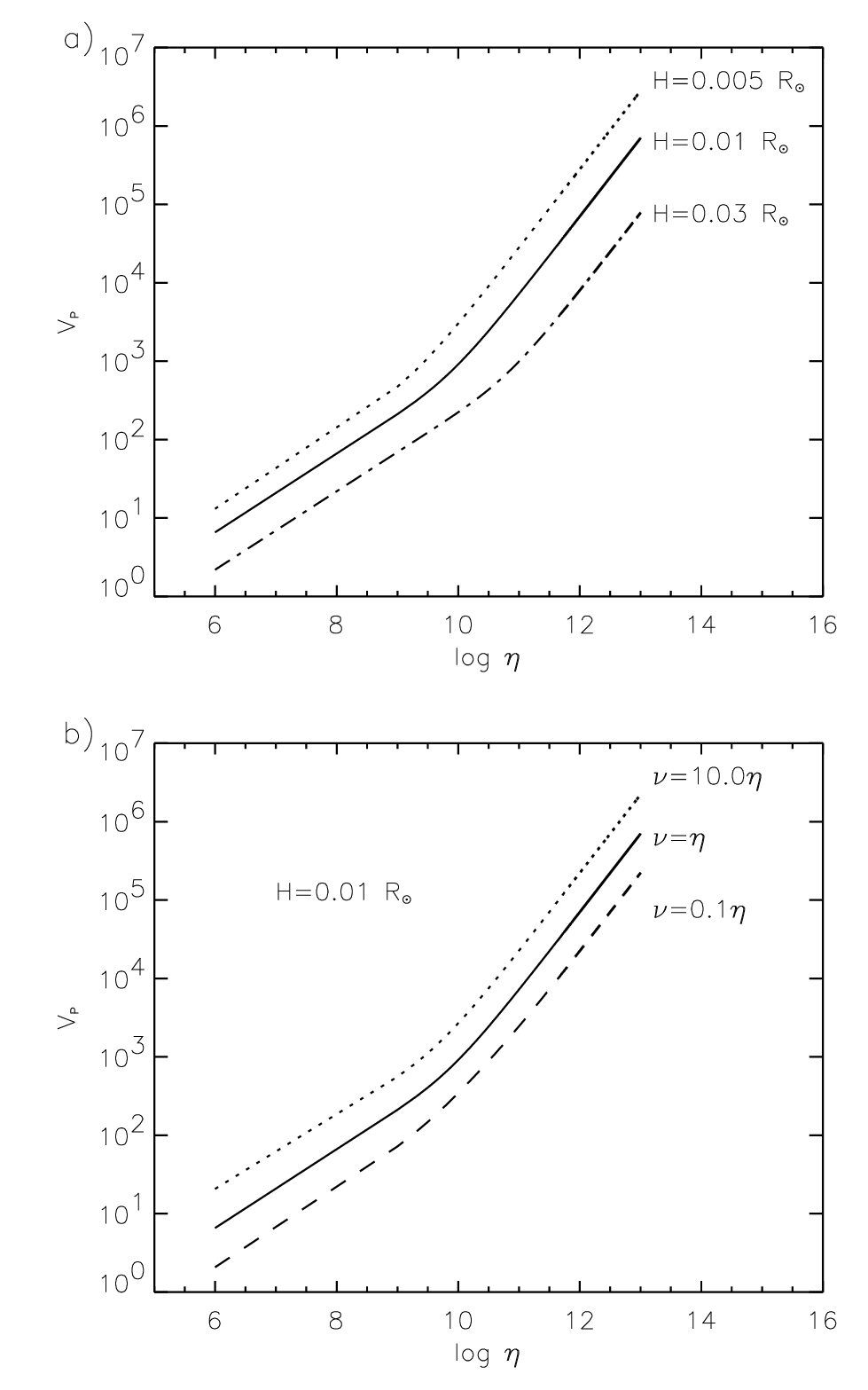}
\caption{Magnetic field strength in Alfv\'en speed units $V_p$ necessary to
confine the tachocline to thickness $H$ as a function of turbulent magnetic 
diffusivity $\eta$: {\it (a)} for $\nu/\eta=1$ with different values of $H$;
{\it (b)} for $H=0.01 R_\odot$ with different values of the Prandtl number.
(Note that, by coincidence, in the solar tachocline $B_p\sim V_p$ to order of
magnitude, in CGS units.)}
\label{fig:estim}
\end{figure}

For an estimate, we will suppose $\ov{\cos(\omega t)f(\omega t+\phi)}\in{\cal
O}(1)$ (i.e.\ no ``conspiracy'' between the phases, a rather natural
assumption). As $H\ll R$ we may approximate $\nabla^2\sim H^{-2}$. Estimating
the other derivatives as $\partial_t\sim\omega$ and $\partial_x\sim R^{-1}$,
(\ref{eq:vmean}) yields
\[  V_p b'/R \sim \nu \ov v/H^2   \label{eq:vmeanest}  \]
A similar order-of-magnitude estimate of the terms in equation (\ref{eq:vfluc})
yields
\[ (\omega +\nu/H^2)v' \sim V_pb'/R   \]
while from (\ref{eq:beq}) we find in a similar manner
\[ \omega b'\sim (V_p+v')V_p/R +\eta b'/H^2   \]
From the last three order-of-magnitude relations one can work out with some
algebra
\[ {V_p^2}= \frac{\nu R^2\omega}{H^2} 
        \frac{(1+\eta/\omega H^2)(1+\nu/\omega H^2)}{1+2\nu/\omega H^2}  
    \label{eq:estimate} \]

Equation (\ref{eq:estimate}) then tells us the field amplitude (in Alfven
units) $V_p$ needed to confine the tachocline to a thickness $H$ with given
values of the diffusivities and of the oscillation period. If the diffusivities
are of turbulent origin one expects $\nu/\eta\simeq 1$ and equation
(\ref{eq:estimate}) can be used to plot $V_p$ as a function of $\eta$ for
different values of $H$, with $\omega=2\pi/22\,$yrs (Fig.~\ref{fig:estim}a).
Figure~\ref{fig:estim}b illustrates the sensitivity of this result to our
assumption about the magnetic Prandtl number $\nu/\eta$. 

Of course, the $(V_p,\eta)$ pairs that reproduce the observed relation $H\simeq
5\,$Mm are also subject to the condition $H\la \Hskin$, otherwise the
assumption of an oscillatory field pervading the tachocline would not be
consistent. Using equation (\ref{eq:skin}), this implies that only the regime
$\eta\ga 10^{9}\,$cm$^2/$s should be considered.

Figure~\ref{fig:estim} thus suggests that an oscillatory poloidal field of
about a thousand gauss is able to confine the tachocline to its observed
thickness for $\eta\sim 10^{10}\,$cm$^2/$s; for higher diffusivities a somewhat
stronger field is needed. This figure can then serve as our guide in choosing
the parameters of the more realistic spherical models considered in the next
section.

\section{Numerical Solution}

\subsection{Equations}

In this section we focus on the numerical solution in the case of spherical
geometry, using the realistic solar stratification. The time
evolution of the velocity  field $\mathbf{v}$ and magnetic field $\mathbf{B}$
are governed by the Navier-Stokes and induction equations:
\begin{equation}
\partial_t \mathbf{v} + \left( \mathbf{v} \cdot \nabla \right) \mathbf{v} = -\nabla V - \frac{1}{\rho} \nabla \left(
p+ \frac{B^2}{8\pi} \right) + \frac{1}{4\pi\rho}(\mathbf{B} \cdot \nabla)\mathbf{B} + \frac{1}{\rho}
\nabla\cdot \mathbf{\tau} , 
\end{equation}
and
\begin{equation}
\partial_t \mathbf{B} = \nabla \times ( \mathbf{v} \times \mathbf{B} ) - \nabla \times ( \eta
\nabla \times \mathbf{B}) ,
\end{equation}
where $\tau$ is the viscous stress tensor, $V$ is the
gravitational potential and $p$ is pressure. These equations are supplemented by the
constraints of mass and magnetic flux conservation
\begin{equation}
\nabla\cdot ( \rho \mathbf{v}) = 0 ,
\end{equation}
\begin{equation}
\nabla \cdot\mathbf{B} = 0 .
\end{equation}
In our model we use the azimuthal component of the equations assuming axial
symmetry and we ignore the meridional flow. Then the magnetic and velocity
fields can be written as
\begin{equation}
\mathbf{B}=\left[ \frac{1}{r^2 \sin\theta}\partial_{\theta} A, \, 
-\frac{1}{r \sin\theta}
\partial_r A, \, B \right], 
\end{equation}
\begin{equation}
\mathbf{v}=r \sin\theta\, \omega(r,\theta,t) \mathbf{e}_{\phi} ,
\end{equation}
where the usual spherical coordinates are used, $A$ is the poloidal field
potential, $B$ is the toroidal field, $\omega$ is the angular velocity and
$\mathbf{e}_{\phi}$ is the azimuthal unit vector (cf.\
\citeNP{Rudiger+Kichat:thin.tacho}). In order to present more transparent
equations  we write the poloidal field potential in the following form:
\begin{equation}
A = a(r,t) \sin^2\theta .
\end{equation}
%where $a$ is a function of $r$ and $t$ only.
The components of the viscous stress tensor appearing in the azimuthal
component of the Navier-Stokes equation are
\begin{eqnarray}
  \tau_{\theta \phi} &=& \tau_{\phi \theta} = \rho \nu \frac{\sin\theta}{r}\, 
     \partial_\theta \left( \frac{v_{\phi}}{\sin\theta} \right) ,\\ 
  \tau_{\phi r} &=& \tau_{r \phi} =  \rho \nu r \,\partial_r \left( 
     \frac{v_{\phi}}{r} \right),
\end{eqnarray}
where $\nu$ is the viscosity.

Thus, the equations, including the effects of diffusion, toroidal field
production by the differential rotation and the Lorentz force, read

\begin{eqnarray}
 \lefteqn{\partial_t \omega = \left( \partial_r \nu + 4 \frac{\nu}{r} + 
  \frac{\nu}{\rho} \partial_r \rho
  \right) \partial_r \omega + \nu \partial^2_r \omega 
  + \frac{3 \nu \cos\theta}{ r^2 \sin\theta} \partial_{\theta} \omega 
  + \frac{\nu}{r^2} \partial^2_{\theta} \omega} \\
&& + \frac{a \cos\theta}{2 \pi \rho r^3 \sin\theta} \partial_r B 
- \frac{\partial_r a}{4 \pi \rho r^3} \partial_{\theta} B 
+ \left( \frac{a \cos\theta}{2 \pi \rho r^4 \sin\theta} \
- \frac{\partial_r a \cos\theta}{4 \pi \rho r^3 \sin\theta} \right) B , \nonumber \\
&& \nonumber \\ 
\lefteqn{\partial_t B = \frac{2 a \sin\theta \cos\theta}{r} \partial_r \omega 
- \frac{\partial_r a \sin^2\theta}{r} \partial_{\theta} \omega} \\
&& + \left( \frac{2 \eta}{r} + \partial_r \eta \right) \partial_r B
+ \eta \partial^2_r B
+ \frac{\eta \cos\theta}{r^2 \sin\theta} \partial_{\theta} B
+ \frac{\eta}{r^2} \partial^2_{\theta} B  \nonumber \\
&& + \left( \frac{\partial_r \eta}{r} - \frac{\eta \cos^2\theta}{r^2 \sin^2\theta} - \frac{\eta}{r^2}
\right) B .\nonumber
\end{eqnarray}
Under the assumption of no meridional circulation used here, these equations
remain unchanged when written in a reference frame rotating with an angular
frequency $\Omega$.

\subsection{Boundary conditions}

The computational domain for the present calculations consists of just the
upper part of the radiative interior, between radii $\rin$ and $\rcz$. We use 
the same boundary conditions for 
$\omega(r,\theta,t)$ as those in \citeN{Elliott}.  We suppose that the rotation
rate at the base of convection zone can be described with the  same expression
as in the upper part of the convection zone. In accordance with the
observations of the GONG network, the following expression is given for
$\Omega_{\mbox{\scriptsize{bcz}}}$:
\begin{equation}
\frac{\Omega_{\mbox{\scriptsize{bcz}}}}{2\pi} = 456 - 72 \cos^2\theta - 42 \cos^4\theta \hspace{0.2cm} \mbox{nHz}.
\end{equation}
This is used to give the outer boundary condition on $\omega$,
\begin{equation}
\Omega + \omega = \Omega_{\mbox{\scriptsize{bcz}}} \hspace{0.5cm} \mbox{at} \hspace{0.1cm}r = r_{\mbox{\scriptsize{bcz}}}.
\end{equation}
$\Omega$ is chosen as the rotation rate in the radiative interior below the
tachocline. On the basis of helioseismic measurements, this value is equal to
the rotation rate of the convection zone at a latitude of about $30^{\circ}$,
corresponding to $\Omega/2\pi \approx 437$ nHz.

The second boundary condition $\omega = 0 $
is imposed at the inner edge of our domain $\rin$.

Our boundary conditions for the toroidal field are simply
\begin{equation}
B=0  \hspace{0.5cm} \mbox{at} \hspace{0.1cm}r = r_{\mbox{\scriptsize{bcz}}} 
\mbox{ and } r = r_{\mbox{\scriptsize{in}}} .
\end{equation}

\subsection{Initial conditions}

The initial conditions chosen for all calculations are
\begin{eqnarray}
\omega (r,\theta , t=0) & = & \Omega_{\mbox{\scriptsize{bcz}}} - \Omega \hspace{0.5cm} \mbox{at}
	\hspace{0.1cm}r = r_{\mbox{\scriptsize{bcz}}} \nonumber\\
\omega (r,\theta , t=0) & = & 0  \hspace{1.8cm} \mbox{at} 
	\hspace{0.1cm}r < r_{\mbox{\scriptsize{bcz}}}\\
B (r,\theta , t=0) & = & 0 . \nonumber
\end{eqnarray}

\subsection{Numerical method}

We used a time relaxation method with a finite difference scheme first order
accurate in time to solve the equations. A uniformly spaced grid, with spacings
$\Delta r$ and  $\Delta \theta$ is set up with equal numbers of points in the
$r$ and $\theta$ directions. $r$ is chosen to vary from $\rin=4.2 \times
10^{10}$ cm to $r_{\mbox{\scriptsize{bcz}}}$, and $\theta$ varies from $0$ to
$\pi/2$.

The stability of this explicit time evolution scheme is determined
by the condition
\begin{equation}
\Delta t< \mbox{Min\,}\{ \Delta t_{\mbox{\scriptsize{diff}}} , \,
\Delta t_{\mbox{\scriptsize{magdiff}}}, \, 
\Delta t_{\mbox{\scriptsize{Maxwell}}} \}
\end{equation}
where
\begin{equation}
\Delta t_{\mbox{\scriptsize{diff}}} = \frac{\Delta r^2}{2\nu_{\mbox{\scriptsize{max}}}}  
\hspace{0.6cm}
\Delta t_{\mbox{\scriptsize{magdiff}}} = \frac{\Delta r^2}{2\eta_{\mbox{\scriptsize{max}}}}  
\hspace{0.6cm}
\Delta t_{\mbox{\scriptsize{Maxwell}}} = \left(\frac\eta\nu\right)^{1/2} 
   \frac{\Delta r}{B_p}  .
\label{eq:idolepes}
\end{equation}
Here, 
%$t_{\mbox{\scriptsize{diff}}}$ is the time step for the diffusion, 
%$t_{\mbox{\scriptsize{magdiff}}}$ is the time step for the magnetic diffusion,
%$t_{\mbox{\scriptsize{Maxwell}}}$ is the time step originating from the Maxwell term, and
$B_p$ is the amplitude of the poloidal field, defined here as
\begin{equation}
  B_p^2 = \mbox{Max\,}\left\{\left| \frac{4 a^2}{\rcz^4} 
  + \frac{(\partial_r a)^2}{\rcz^2} \right| \right\} .
\end{equation}

Our calculations are based on a more recent version of the solar model of 
\citeN{Guenther}.

\begin{figure}[t]
\centering
\includegraphics[width=8cm]{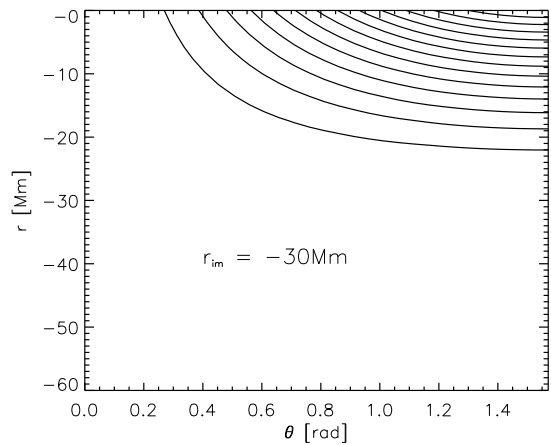}
\caption{The poloidal field configuration.}\label{fig:vectpot}
\end{figure}

\section{Numerical Results and Discussion}

In this section we examine the influence of an oscillatory magnetic field on
the radial spreading of the differential rotation on the basis of our numerical
results. In our calculations the poloidal magnetic field is assumed to be time
independent in amplitude, oscillating in time and {\it a priori} known. For the function $a$ of the poloidal 
field potential the  following expression is used:
\begin{eqnarray}
a & = & A_0 \cos(\omega t) \frac{r^2_{\mbox{\scriptsize{bcz}}}}{2} 
%KRIS:   \left( \frac{r-r_{\mbox{\scriptsize{im}}}}{\rcz-\rin} \right)^2
%EM: ez javitas, eredetileg ez volt (remelhetoleg csak eliras):
\left( \frac{r-r_{\mbox{\scriptsize{im}}}}{r_{\mbox{\scriptsize{bcz}}}-r} \right)^2
\hspace{0.6cm}
r \ge r_{\mbox{\scriptsize{im}}} \\
a & = & 0
\hspace{5.0cm}
r_{\mbox{\scriptsize{im}}} > r, \nonumber
\end{eqnarray}
where $A_0$ fixes the field amplitude and $r_{\mbox{\scriptsize{im}}}$ is the
depth of the penetration of the poloidal magnetic field into the radiative
interior. As the skin effect should limit the penetration of the oscillatory
field below the turbulent tachocline, we set $\rim=30\,$Mm. Figure
\ref{fig:vectpot} illustrates the poloidal field geometry. Note that the radial
coordinate $r$ shown on the ordinates in our figures has its zero point reset
to $\rcz$, i.e.\ $r$ in the figures corresponds to $r-\rcz$.  Thus, the
negative $r$ values correspond to the layers below the convective zone.

The diffusive timescale over which the solution should relax to a very nearly
periodic behaviour is
\begin{equation}
\tau={(r_{\mbox{\scriptsize{bcz}}}-r_{\mbox{\scriptsize{in}}})^2}/{\eta_{\mbox{\scri min}}}
\end{equation}
where $\eta_{\mbox{\scri min}}$ is the lowest value of $\eta$ in the domain,
i.e.\ its value taken at $\rin$. Physically we would expect this value to be
close to the molecular viscosity. Using this value in the computations would,
however, lead to a prohibitively high number of timesteps to relaxation.
(Clearly, the runtime of  our computations must be chosen to well exceed $\tau$
to reach relaxation.) 

\begin{figure}[!t]
\centering
\includegraphics{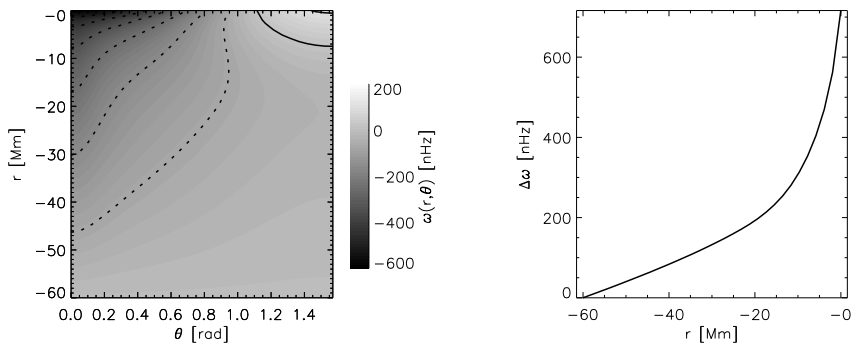}
\caption{Spreading of the differential rotation into the radiative interior 
for $B_p=1603\,$G.
{\it Left-hand panel:} contours of the time-average of the angular rotation rate
$\omega(r,\theta,t)$  under one dynamo period. Equidistant contour levels are
shown, separated by intervals of $100\,$nHz, starting from 0 towards both
non-negative (solid) and negative (dashed) values. {\it Right-hand panel:} differential
rotation amplitude $\Delta\omega$ (defined as the difference of maximal and
minimal $\omega$ values in a horizontal surface) as a function of radius.}
\label{fig:a0_100}
\end{figure}

\begin{figure}[!h]
\centering
\includegraphics{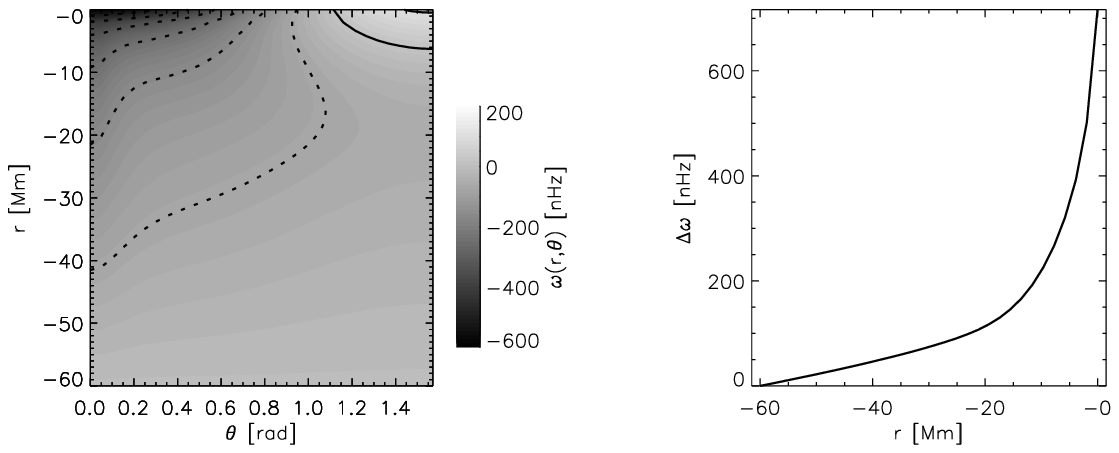}
\caption{Same as in Figure~\ref{fig:a0_100} for $B_p=2004\,$G}
\label{fig:a0_125}
\end{figure}

\begin{figure}[!pt]
\centering
\includegraphics{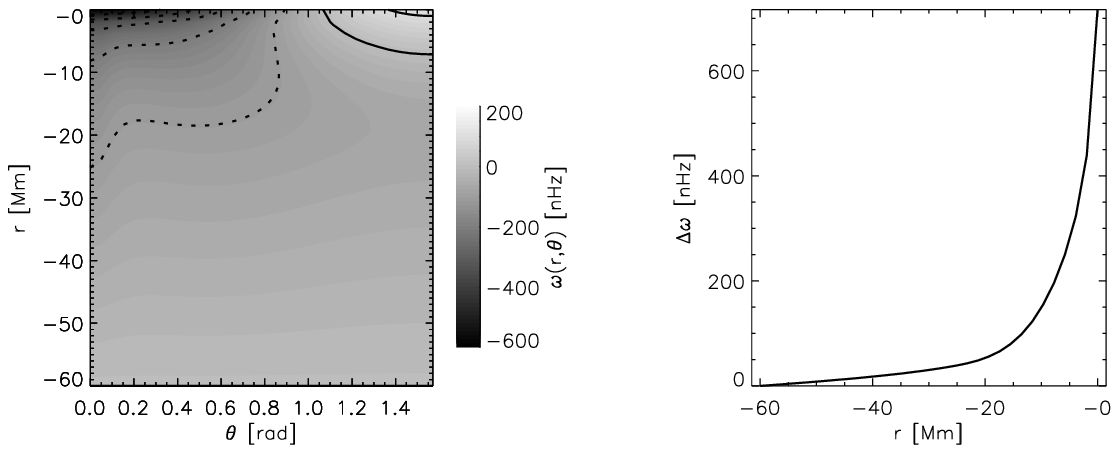}
\caption{Same as in Figure~\ref{fig:a0_100} for $B_p=2405\,$G}
\label{fig:a0_150}
\end{figure}

Therefore, we first consider the simpler case 
$\eta=10^{10}\,$cm$^2/$s$\,=\,$const.,  corresponding to $\tau\sim
120\,$years.  The results for this case are shown in 
Figure~\ref{fig:a0_100}--\ref{fig:a0_150} for  different amplitudes of the
poloidal magnetic field, after relaxation. In accordance with the results of
Section~2 it is found that a kilogauss poloidal field (peak amplitude 2400 G)
is able to confine the tachocline to its observed thickness.

\begin{figure}[!h]
\centering
\includegraphics[width=8cm]{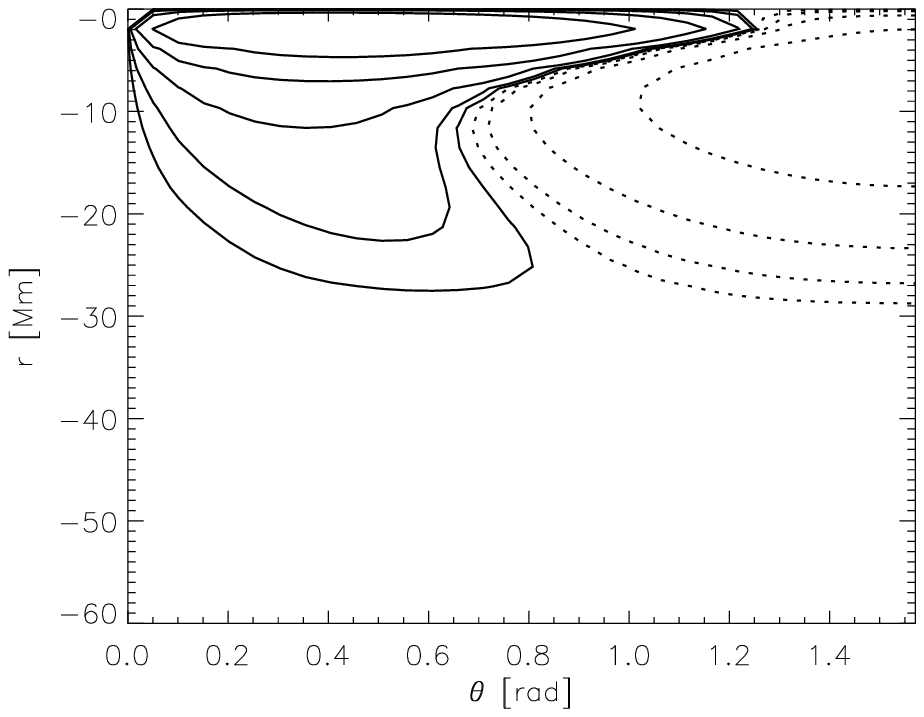}
\caption{Contours of the time-averaged azimuthal component of the  
Lorentz acceleration $f_L(r,\theta,t)$ for the case in Fig.~\ref{fig:a0_150}. Contour levels correspond
to $\log |f_L|=(-6,-5.5,-5,-4.5,-4)$, solid contours standing for $f_L>0$,
dashed contours for $f_L<0$.}
\label{fig:Lorentz}
\end{figure}

\begin{figure}[!pt]
\centering
\includegraphics[width=11.8cm]{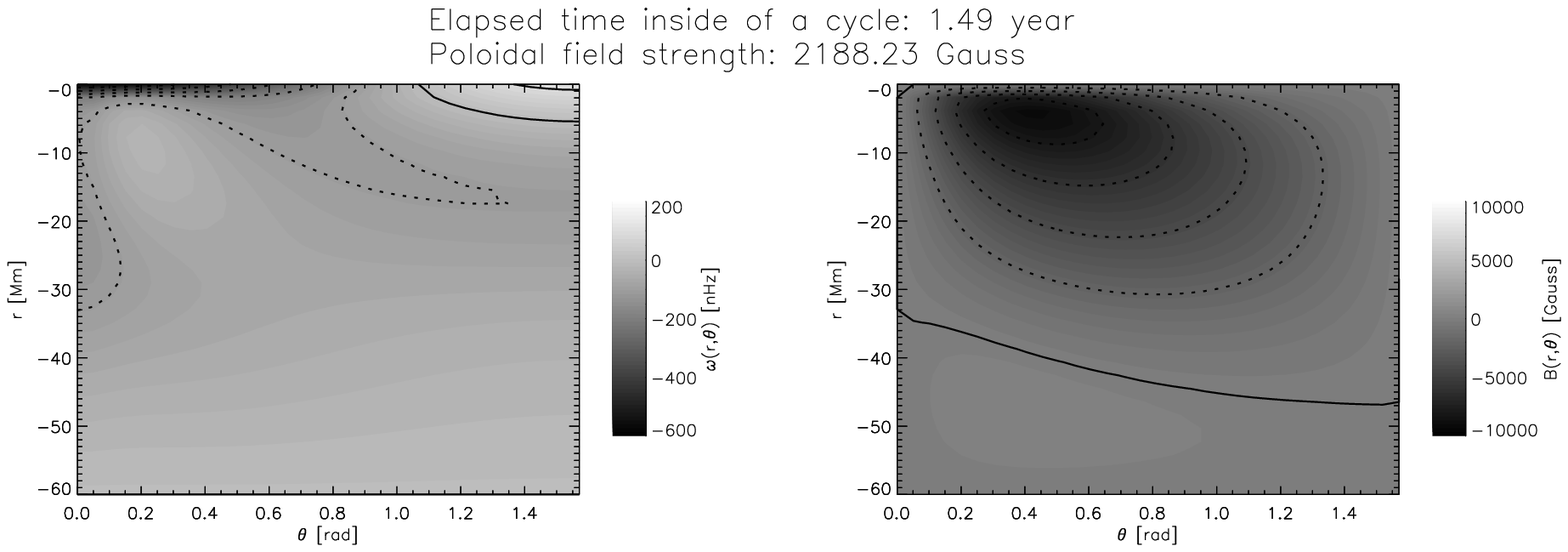}\\
\includegraphics[width=11.8cm]{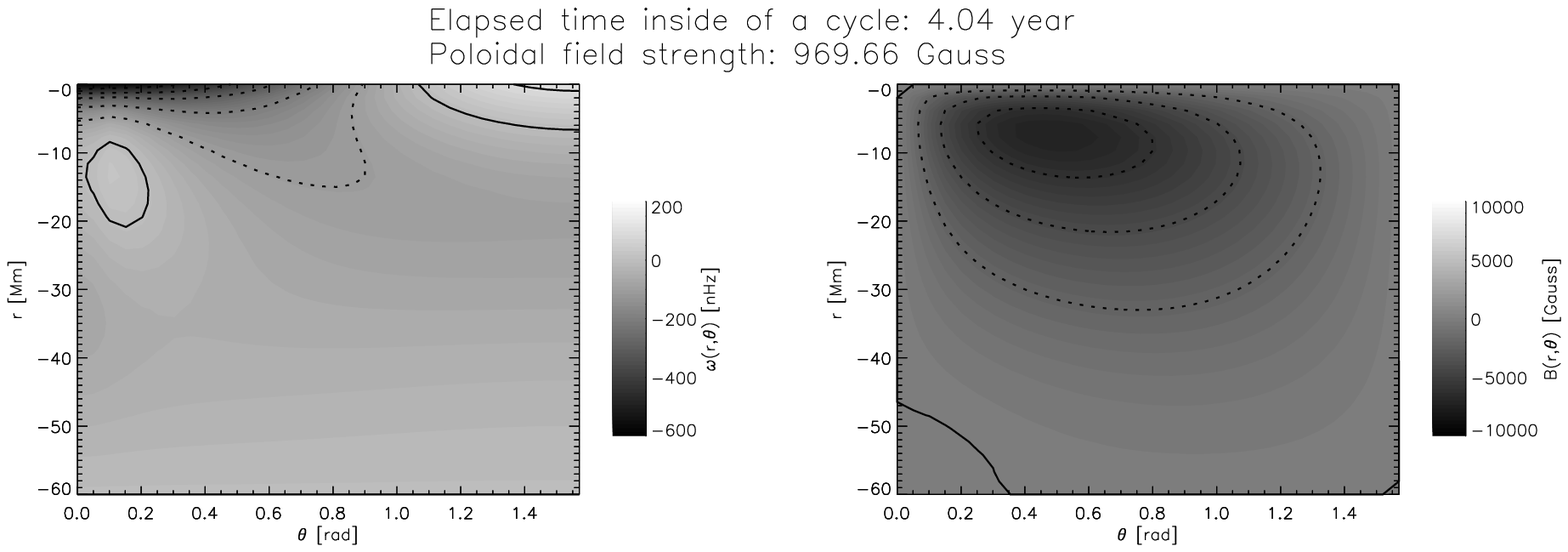}\\
\includegraphics[width=11.8cm]{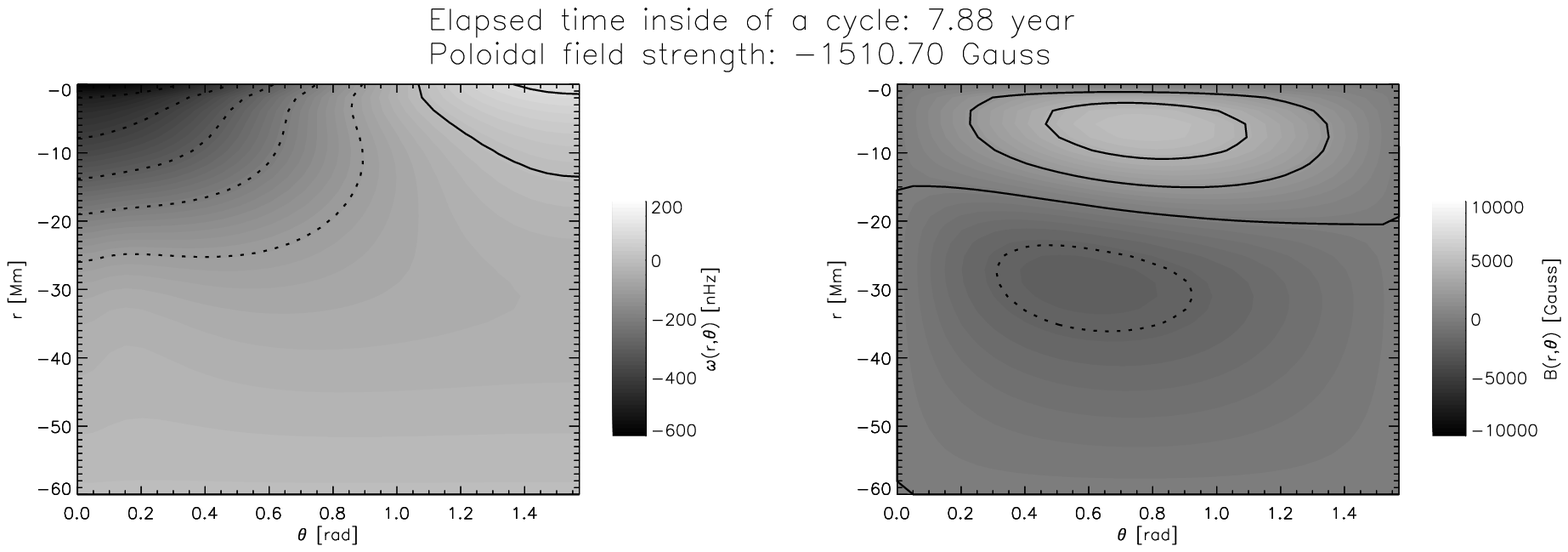}\\
\includegraphics[width=11.8cm]{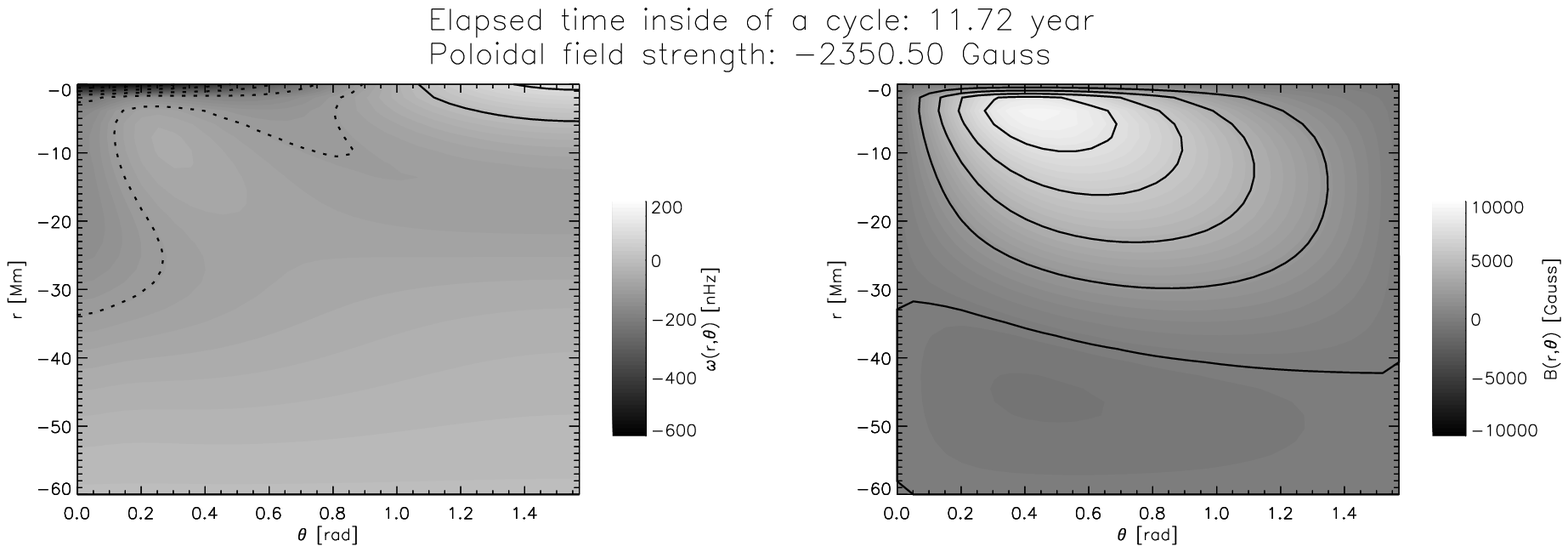}\\
\caption{Snapshot of the solution at four cycle phases for $B_p=2405\,$G. $t=0$
corresponds to poloidal field maximum. 
{\it Left-hand panels: } contours of angular rotation rate, as in Fig.~\ref{fig:a0_100}. 
{\it Right-hand panels: }
contours of toroidal magnetic field strength. Equidistant contour levels are
shown, separated by intervals of $2000\,$G.}
\label{fig:fejlodes}
\end{figure}

Figure~\ref{fig:Lorentz} presents the distribution of the azimuthal component of the
 Lorentz acceleration 
%%\[ f_L=(\vc j\times\vc B)_\phi/c\rho \]
\[ f_L=\frac{1}{4\pi\rho}(\mathbf{B} \cdot \nabla) B_\phi \]
in a meridional section of the tachocline, showing polar acceleration and equatorial 
deceleration.
The time variation of the differential rotation and the toroidal magnetic field
during a cycle after relaxation is shown in Fig.~\ref{fig:fejlodes}. Note the
significant variation with cycle phase. 

\begin{figure}[!t]
\centering
\includegraphics[width=8cm]{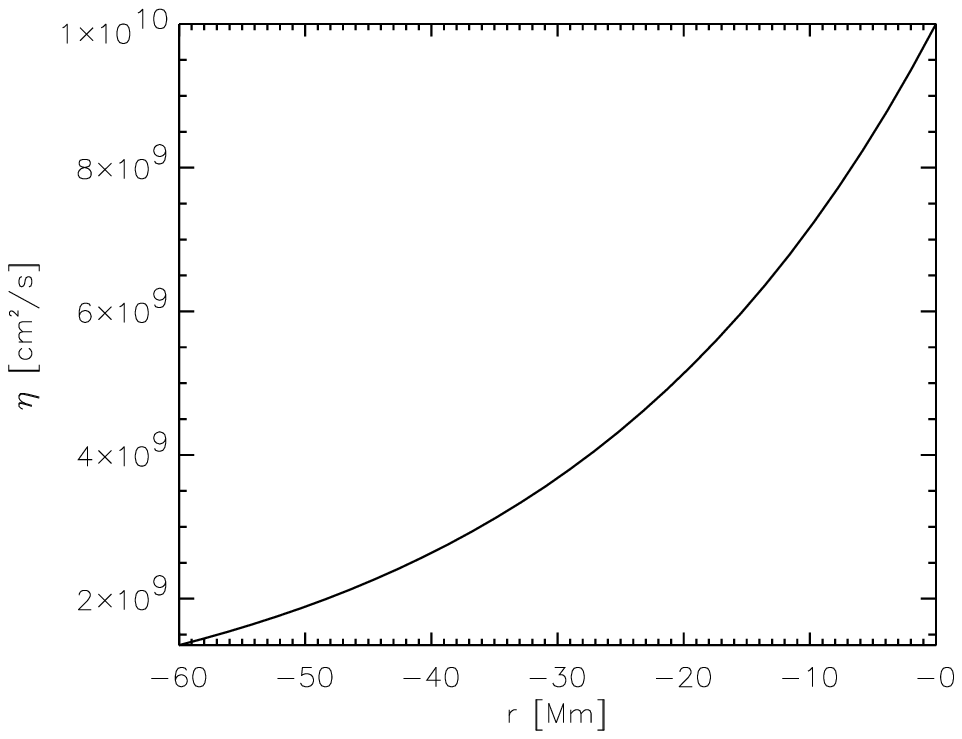}
\caption{Radial profile of the diffusivity used for Figure~\ref{fig:a0_150_var}.}\label{fig:varnu}
\end{figure}

\begin{figure}[!h]
\centering
\includegraphics{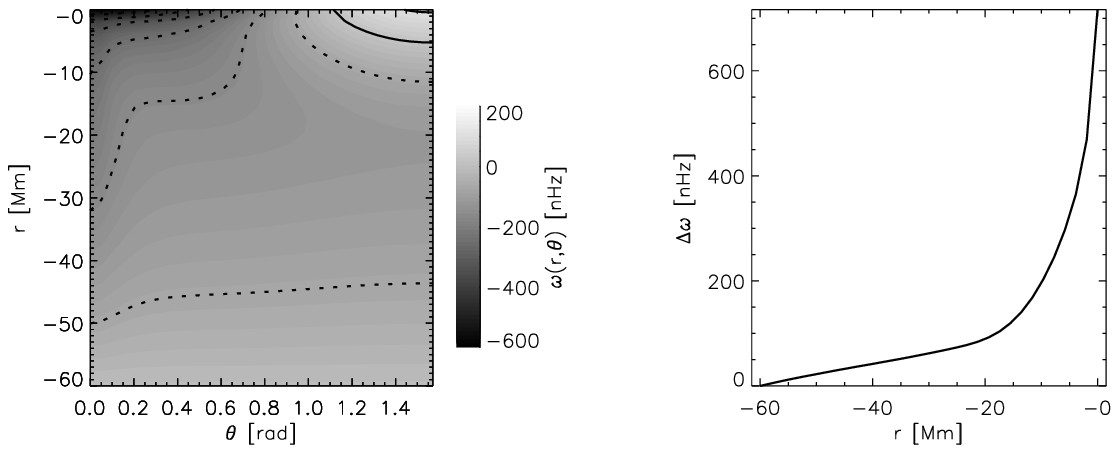}
\caption{The same as in Figure~\ref{fig:a0_150}, but the diffusivity is varied 
in radius.}\label{fig:a0_150_var}
\end{figure}

Finally, in the calculation presented in 
Figure~\ref{fig:varnu}--\ref{fig:a0_150_var} the diffusivity is 
allowed to vary with radius  as
\begin{equation}
\eta=\eta_0 \exp \left( 
\frac{r-r_{\mbox{\scriptsize{bcz}}}}{r_{\mbox{\scriptsize{bcz}}}-r_{\mbox{\scriptsize{im}}}} \right).
\end{equation}
This variation, while quantitatively much milder than expected physically,
allows us to consider the effects of a variable diffusivity without the
necessity of prohibitively long integration times. The influence of the
downwards decreasing diffusivity on the solution is found to be small.

Now we turn to the comparison of the detailed spatio-temporal behaviour of our
solutions to other models and helioseismic measurements. As a caveat we should
stress that, in contrast to the above conclusions about tachocline confinement,
these details of our models are unlikely to represent the situation in the real
Sun faithfully, as the assumption of a simple dipole geometry (i.e. the use of
a ``standing wave'' instead of a travelling dynamo wave) is clearly not
realistic. Yet it may be interesting from a theoretical point of view to draw
some parallels with other work. In particular, the polar ``cone'' of rotational
deceleration discernible in Figs.~\ref{fig:a0_100}--\ref{fig:fejlodes} is
reminiscent of the similar feature seen in models with a steady internal field
(cf.~Fig.~2C of \citeNP{McGregor+Char}). This feature is, however, likely to
disappear or change completely in a model with a poloidal field prescribed as a
travelling dynamo wave.

\begin{figure}[!t]
\centering
\noindent
\begin{minipage}{100mm}
\resizebox{100mm}{!}{
\rotatebox{-90}{
\includegraphics{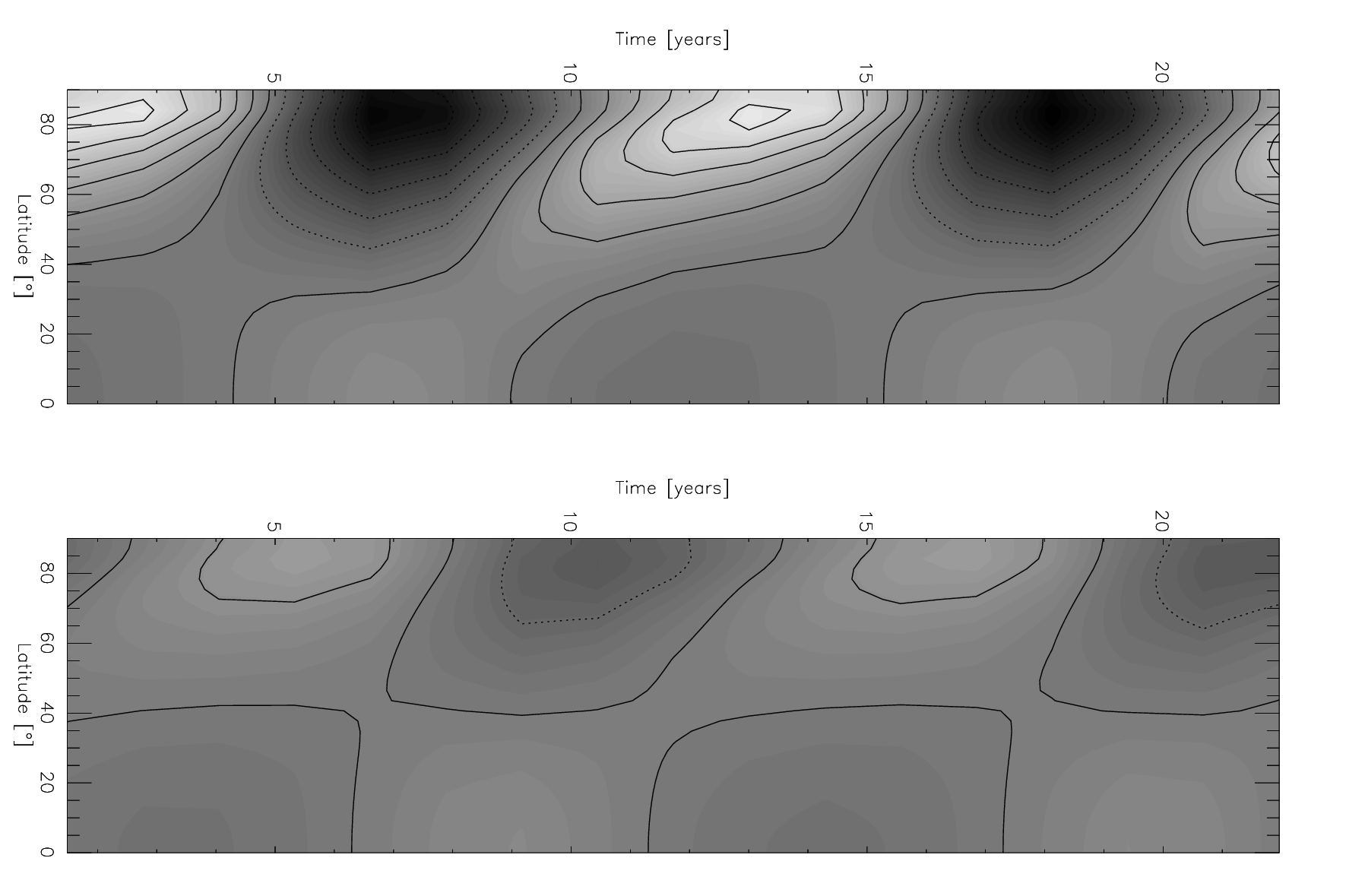}}}
\end{minipage}
\begin{minipage}{10mm}
\resizebox{10mm}{!}{
\includegraphics{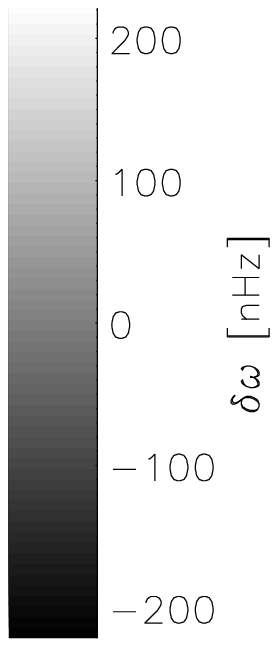}}
\end{minipage}
\caption{Time--latitude diagram of fluctuations of the angular velocity $\omega$
around its temporal average at a given latitude for the case in Fig.~\ref{fig:a0_150}, 
at two different levels in the
tachocline. {\it Top: } $r-\rcz=-3.87$ Mm; {\it Bottom: } $r-\rcz=-27.09$ Mm}
\label{fig:migr}
\end{figure}

In Fig.~\ref{fig:migr} we present the time-latitude diagram of angular velocity
fluctuations during one cycle. What is remarkable in this plot is not so much
the weak poleward migration (which is sensitive to model details, and for other
forms of $A(r,\theta,t)$ completely different migration patterns may result)
but the fact that no other periodicity than the imposed 11/22 years is visible.
In particular, there is no trace of the controversial 1.3 year periodicity
recently claimed by several authors (\citeNP{Corbard+:SOGO} and references
therein).

Computer animations of the time evolution illustrated in Fig.~\ref{fig:fejlodes}
and other cases may be found on the accompanying CD-ROM, or they can be downloaded from the following internet 
address:\\
{\tt http://astro.elte.hu/kutat/sol/mgconf/mgconfe.html}.

\section{Conclusion}

We have demonstrated that the tachocline can be confined to its observed
thickness by a dipole-like poloidal magnetic field of about one kilogauss,
penetrating below the convective zone and oscillating with a period of 22
years, if the tachocline region is turbulent with a diffusivity of $\eta\sim
10^{10}\,$cm$^2/$s. Our estimates presented in Section~2 suggest that a similar
confinement may be produced for other pairs of the parameter values ($B_p$,
$\eta$). For the numerical computations we assumed a magnetic Prandtl number
$\nu/\eta=1$; Fig.~\ref{fig:estim}(b) gives an impression of the influence of
using another $\nu/\eta$ value on the quantitative results

A crucial assumption in these calculations is that the oscillatory poloidal 
magnetic field does penetrate below the convective zone to depths comparable to
the tachocline thickness. This remains an assumption here, the penetration
being prescribed a priori by setting the parameter
$r_{\mbox{\scriptsize{im}}}$. While this should be confirmed later in more
general calculations including the evolution equation for the poloidal field,
formula (\ref{eq:skin}) does suggest that the assumption is consistent whenever
$\eta\ga 10^9\,$cm$^2/$s. 

This explains the necessity of the high turbulent diffusivity in the
tachocline, used in the present models. The high diffusivity, in turn, is the
principal reason why the field strength necessary for tachocline confinement is
so much higher here than in the case of models with a steady internal remnant
magnetic field  (\citeNP{McGregor+Char}; \citeNP{Rudiger+Kichat:thin.tacho}; 
\citeNP{Garaud:SOGO}): as the turbulent magnetic Prandtl number is not expected
to differ greatly (i.e.\ by more than an order of magnitude) from unity,
viscous angular momentum transfer is also quite effective. 

Another effect that may contribute to the high field strength needed is that
here the field is fully anchored in the differentially rotating overlying
envelope: since $B_r$ is nonzero at the upper boundary, the magnetic field
applies stresses vertically as well as horizontally. In ideal MHD, for a
stationary state, magnetic stresses are known to impose $\omega=\,$const. along
field lines (Ferraro's theorem).  For finite but low diffusivities, field line
anchoring can still make a great difference in the resulting rotational
profile, as demonstrated by \citeN{McGregor+Char}. The fact that periodic
oscillations of $\omega$ around its temporal mean in our model
(Fig.~\ref{fig:migr}) are strongest near the poles, where the field is nearly
radial, shows that, despite the high diffusivity, this effect is to some extent
also significant in our models. Thus, field line anchoring may be responsible
for the fact that the field strengths required in the numerical models exceed
somewhat the analytic estimates of Section~2, based on the assumption of a
horizontal field.

The $\sim 1000\,$G poloidal field strength assumed here is nonetheless not
implausible. Indeed, transport equilibrium models of the poloidal field inside
the convective zone (\citeNP{Petrovay+Szakaly:2d.pol}) show that the
characteristic strength of the poloidal field in the bulk of the convective
zone is order of $10\,$G; flux conservation then sets a lower limit of a few
hundred gauss for the poloidal field strength in the region below the
convective zone where the field lines close. This is quite compatible with peak
values of 1-2000 G.

Our neglect of meridional circulation may also seem worrying to some, given
that \citeN{Spiegel+Zahn} found that the Eddington--Sweet circulation
transports angular momentum much more efficiently than viscosity, and so it is
the primary mechanism responsible for the inward spreading of the tachocline.
This is, however, not expected to be the case for a turbulent tachocline. Owing
to the strongly subadiabatic stratification below the convective zone, the
timescale of any meridional circulation cannot be shorter than the relevant
thermal diffusive timescale (to allow moving fluid elements to get rid of their
buoyancy). In the present case the relevant heat conductivity is the turbulent
one, so for turbulent Prandtl numbers not very different from unity, the
shortest possible timescale for meridional circulation is just comparable to
the viscous timescale. Thus, while meridional circulation may possibly indeed
have a significant effect on tachocline structure, in our view it is not likely
to dominate over turbulent viscosity.

%Our results suggest that a plausible confinement mechanism for the tachocline
%is angular momentum transport by Maxwell stresses in a dynamo-generated,
%oscillatory magnetic field. As the strong shear present in the tachocline is
%an essential ingredient of many dynamo models, this points to the interesting
%possibility that the dynamo itself maintains the shear necessary for its
%operation. Coupled with the increasingly widespread notion that the
%$\alpha$-effect may also be based on strong fields generated by the dynamo,
%the possibility arises that the dynamo operates in a ``homeostatic'' mode,
%maintaining the conditions necessary for its own existence.

It is clear that the models presented in this paper are still far from being a
realistic representation of the solar tachocline. In particular
\begin{lista}
\item In order to limit the number of free parameters, we made no attempt here
to  simulate the latitudinal migration of the magnetic field during the cycle.
\item The diffusivity, if generated by instabilities of the tachocline, should
in some way be related to (e.g.) the local shear, instead of being arbitrarily
specified.
\item The poloidal field evolution should also be consistently calculated in a
more complete model.
\item Meridional circulation may also be taken into account.
\end{lista}
And the list could be continued. Despite these simplifications, we believe that
our model convincingly demonstrates the feasibility of tachocline confinement
by a dynamo field. Possible generalizations of these models can be made
according to the remarks listed above. Work in this direction is in progress.

\acknowledgements 
We thank D. B. Guenther for making his solar model available. This work was 
funded by the OTKA under grants no.\ T032462 and T034998.

%\bibliography{kris}
%\bibliographystyle{solphys}

\begin{ao}
\\
E. Forg\'acs-Dajka\\
E\"otv\"os University, Dept.~of Astronomy\\
Budapest, Pf.~32, H-1518 Hungary\\
E-mail: andro@astro.elte.hu
\end{ao}

\end{article}
\end{document}